\documentstyle[12pt]{article}
\begin{document}
\begin{titlepage}
\hfill \vbox{
    \halign{#\hfil         \cr
           IPM/P-2002/007 \cr
           hep-th/0204117 \cr
           } 
      }  
\vspace*{20mm}
\begin{center}
{\Large {\bf }Gravity on noncommutative D-branes \\ }

\vspace*{15mm} \vspace*{1mm} {F.
Ardalan,${}^{a,b,}$\footnote{Ardalan@theory.ipm.ac.ir} H.
Arfaei,${}^{a,b,}$\footnote{Arfaei@theory.ipm.ac.ir} M. R.
Garousi$ {}^{a,c,}$\footnote{Garousi@theory.ipm.ac.ir} and A.
Ghodsi${}^{a,b,}$\footnote{Ahmad@theory.ipm.ac.ir}}\\

\vspace*{1cm} {\it $^a$ Institute for Studies in Theoretical
Physics and Mathematics (IPM),\\
P.O. Box 19395-5531, Tehran-Iran\\
$^b$ Department of Physics, Sharif University of
Technology,\\
P.O. Box 11365-9161, Tehran-Iran\\
$^c$ Department of Physics, Ferdowsi University, Mashhad-Iran\\}

\vspace*{1cm}
\end{center}

\begin{abstract}
The effective action for the low energy
scattering of two gravitons with a D-brane
in the presence of a constant antisymmetric
$B$ field in bosonic string theory is
calculated and the modification to the
standard D-brane action to first order in
$\alpha'$ is obtained.
\end{abstract}
\end{titlepage}
\newpage
\section{Introduction}
The discovery in the early days of
string theory that string
amplitudes in low energy regimes maybe
reproduced by Yang-Milles
field theory for open strings and
gravitational field theory for
closed strings, was the beginning of
a long, fruitful study into
the relation of string theory and field
theory in general and in low energy
in particular \cite{r1,r2}.

The lowest order in momentum expansion of the
low energy string theory confirmed the
interpretation of the massless string modes
as the gauge bosons and gravitons of the
respective theories, and the next order was
a prediction for the corrections to the gauge
theory and to the gravity\cite{r3,r4}.

The realization of the significance of
D-branes in string theory, correspondingly
led to the study of the low energy effective
dynamics of these objects in addition to
the bulk dynamics \cite{r5,r6,r7,r8,r9,r10,r15}.
The lowest order in the momentum expansion
turned out to be the DBI action with higher order
corrections involving nontrivial terms,
both in the bosonic string and the superstring theory.

Discovery of noncommutativity in string
theory in the presence of a nontrivial
background \cite{r11} resulted in the dramatic
resurgence of noncommutative gauge theories
with their highly nontrivial novel properties
emerging as deformation of the low energy
action of open strings in the presence of a background.
Higher order corrections to the
noncommutative gauge theory have also been derived.

It is then natural to ask what kind
of deformation the low energy effective
action of closed strings suffers in the
presence of constant back ground antisymmetric field.\\
Noncommutative deformations of gravity
have been proposed and studied recently \cite{r12}.
However they have not been derived from string
theory and are ad hoc in this sense. One of the
motivations of the present study is to throw
light on these deformations.

In this work we begin to address this problem
by looking at the low energy amplitude of two
closed strings with a D-brane in the presence
of a constant background $B$ field. We find
that as expected the bulk action is not
modified, however, the effective action describing
the dynamics of the D-brane is modified due to
the presence of the constant B field on the D-brane.
However the effective action is not obtainable,
as a naive T-duality might indicate, from the action
of vanishing $B$, by replacing the metric $g$ by $g+B$.

This might be understood on the basis of the
fact that T-duality, at this order, mixes
graviton scattering with antisymmetric field
scattering and is highly nontrivial.

In section two we write down the amplitude of
scattering of two gravitons with a D-brane in
the presence of a B field and find its low
energy behaviour. In section three we find
the low energy effective action describing
those amplitude to order $\alpha'$.
Section four is devoted to discussion, in particular on the issue of
covariance of the effective action.
\section{String scattering from D-branes}
In this section we compute the tree level
bosonic string scattering amplitude of two
massless closed string off a noncommutative
D-brane. We take D-brane as a disk and conformaly
mapped on the upper half plane, with mixed
boundary conditions and use the following
notation for indices, on and off the D-brane
\begin{equation}
\mu,\nu=0,..,25\,\,\,\,\,,\,\,\,\,\,
a,b=0,..,p\,\,\,\,\,,\,\,\,\, i,j=p+1,..,25.
\end{equation}
For a D-brane localized at $x^{p+1},...,x^{25}$,
the boundary conditions are
\begin{equation}
\eta_{ab}(\partial-\overline{\partial})X^{a}+B_{ab}
(\partial+\overline{\partial})X^{a}\mid_{z=\bar{z}}=0.
\end{equation}
The two points correlator of string coordinates
$X^{\mu}(z,\bar{z})$ on the D-brane is
\begin{eqnarray}
\langle X^{\mu}_{i} X^{\nu}_{j}\rangle=
{-\alpha'\over2}\bigg[\eta^{\mu\nu}log
((z_{i}-z_{j})(\bar{z}_{i}
-\bar{z}_{j}))+D^{\mu\nu}log(z_{i}-
\bar{z}_{j})+D^{\nu\mu}log(\bar{z}_{i}-z_{j})\bigg]
,\label{cor}
\end{eqnarray}
where
\begin{equation}
D^{ba}=2({1\over\eta+B})^{ab}-\eta^{ab}
\,\,\,\,,\,\,\,\,
D^{ij}=-\delta^{ij}\,\,\,\,,\,\,\,\,
{D^{\mu}}_\alpha D^{\nu\alpha}=\eta^{\mu\nu},
\end{equation}
with $g$ the flat metric. Using these  correlators
we can compute the scattering amplitude
of two closed strings off a D-brane,
\begin{equation}
A=g_{c}^2e^{-\lambda}\int d^2z_1
d^2z_2\langle V(z_1,\bar{z}_1)
V(z_2,\bar{z}_2)\rangle, \label{amp}
\end{equation}
with $g_c$ the closed string coupling
constant and $\lambda$ the Euler number
of the world sheet. The appropriate vertex
operators for the closed strings are,
\begin{equation}
V(z_i,\bar{z}_i)=\epsilon_{\mu\nu}:\partial
X^{\mu}(z_i)\exp(ik_i.X(z_i))::\bar{\partial}
\widetilde{X}^\nu(\bar{z}_i)\exp(i
k_i\widetilde{X}(\bar{z}_i)):,
\end{equation}
which the momenta $k_i$ and polarizations
$\epsilon_{\mu\nu}$ of the massless closed
strings satisfying
\begin{equation}
\epsilon_{\mu\nu}k^\mu=\epsilon_{\mu\nu}k^\nu=0
\,\,\,\,\,,\,\,\,\,\, k^2=0.
\end{equation}
To go further we use the change of variable as follow,
\begin{equation}
\widetilde{X}^\mu(\bar{z}_i)
={D^\mu}_\xi X^\xi(\bar{z}_i),
\end{equation}
so that the closed string vertices change to
\begin{equation}
V(z_i,\bar{z}_i)=\epsilon_{\mu\lambda}
{D^\lambda}_\nu:\partial X^\mu(z_i)\exp(i
k_i.X(z_i))::\bar{\partial}X^\nu(\bar{z}_i)
\exp(ik_i.D.X(\bar{z}_i)):.
\end{equation}
Now inserting these vertices into (\ref{amp})
and going through the calculation of the
correlation functions we find
\begin{equation}
A=g_{c}^2e^{-\lambda}\epsilon_{1\mu\lambda}
{D^\lambda}_\nu\epsilon_{2\alpha\rho}
{D^\rho}_\beta\int d^2z_1 d^2z_2
{\mathcal{A}}^{\mu\nu\alpha\beta},\label{amp1}
\end{equation}
where
\begin{eqnarray}
{\mathcal{A}}^{\mu\nu\alpha\beta} \!\!\! &=& \!\!\!
iC_{D_2}^X (2\pi)^{p+1}\delta^{p+1}(k_1+k_1.D
+k_2+k_2.D)
\exp\bigg\{{\alpha'\over2}\bigg[k_1.D.k_1
log(z_1-\bar{z}_1)\cr &&\cr \!\!\!
&+& \!\!\! k_2.D.k_2 log(z_2-\bar{z}_2)
+k_1.k_2log(z_1-z_2)+k_1.k_2
log(\bar{z}_1-\bar{z}_2)\cr &&\cr
\!\!\! &+& \!\!\! k_2.D.k_1 log(z_1-\bar{z}_2)
+k_1.D.k_2 log(\bar{z}_1-z_2)\bigg]\bigg\}
\cr &&\cr \!\!\! &\times&\!\!\! \bigg\{
q^{\mu\nu}_{1\bar{1}}q^{\alpha\beta}_{2\bar{2}}
+q^{\mu\alpha}_{12}q^{\nu\beta}_{\bar{1}\bar{2}}
+q^{\mu\beta}_{1\bar{2}}q^{\nu\alpha}_{\bar{1}2}
+q^{\mu\nu}_{1\bar{1}}f^{\alpha}_{2}f^{\beta}_{\bar{2}}
+q^{\mu\alpha}_{12}f^{\nu}_{\bar{1}}f^{\beta}_{\bar{2}}
+q^{\mu\beta}_{1\bar{2}}f^{\nu}_{\bar{1}}f^{\alpha}_{2}
\cr &&\cr \!\!\! &+& \!\!\!q^{\nu\alpha}_{\bar{1}2}
f^{\mu}_{1}f^{\beta}_{\bar{2}}+q^{\nu\beta}_
{\bar{1}\bar{2}}f^{\mu}_{1}f^{\alpha}_{2}
+q^{\alpha\beta}_{2\bar{2}}f^{\mu}_{1}f^{\nu}_{\bar{1}}
+f^{\mu}_{1}f^{\nu}_{\bar{1}}f^{\alpha}_{2}
f^{\beta}_{\bar{2}}\bigg\},\label{amp2}
\end{eqnarray}
in which $C_{D_2}^X$ is the functional
determinant for the string fields and the
delta function ensures conservation of
momentum parallel to the D-brane.
Here $f$'s are terms that come from correlation
function between derivative and exponential terms
while the $q$'s are between the derivative terms,
and are
\begin{equation}
q^{\mu\nu}_{ij}=-{\alpha'\over2}{\eta^{\mu\nu}
\over(z_i-z_j)^2} \,\,\,\,\,,\,\,\,\,\,
f^{\mu}_{i}=-i{\alpha'\over2}\sum_{j\neq
i}({k^{\mu}_{j}\over z_i-z_j}),\label{fq}
\end{equation}
with the indices as
\begin{equation}
i,j=1,\overline{1},2,\overline{2} \,\,\,\,,\,\,\,\,
z_{\overline{1}}=\overline{z}_{1}\,\,\,\,,\,\,\,\,
z_{\overline{2}}=\overline{z}_{2}.
\end{equation}
It is straight forward to see that the
amplitude is SL(2,R) invariant.
To fix this invariance we let $z_1=iy$ and $z_2=i$
\begin{equation}
d^2z_1d^2z_2\rightarrow 2(1-y^2)dy.
\end{equation}
The $f$ and $q$ terms then become
\begin{eqnarray}
&& f^{\mu}_1={1\over 2iy}({y+1 \over y-1}
k^{\mu}_2+{y-1 \over y+1}k_2.D^\mu)
\,\,\,\,,\,\,\,\, f^{\nu}_{\bar{1}}={-1\over
2iy}({y-1 \over y+1}k^{\nu}_2+{y+1 \over
y-1}k_2.D^\nu)\cr &&\cr &&
f^{\alpha}_2={-1\over 2i}({y+1 \over y-1}
k^{\alpha}_1+{y-1 \over y+1}k_1.D^\alpha)
\,\,\,\,,\,\,\,\, f^{\beta}_{\bar{2}}={1
\over 2i}({y-1 \over y+1}k^{\beta}_1+{y+1 \over
y-1}k_1.D^\beta)\cr &&\cr
&&q^{\mu\nu}_{1\bar{1}}={-\eta^{\mu\nu}
\over4y^2}\,\,\,\,,\,\,\,\, q^{\mu\alpha}_{12}
={-\eta^{\mu\alpha}\over(y-1)^2}\,\,\,\,,\,\,\,\,
q^{\mu\beta}_{1\bar{2}}={-\eta^{\mu\beta}
\over(y+1)^2}\cr &&\cr &&q^{\nu\alpha}_
{\bar{1}2}={-\eta^{\nu\alpha}\over(y+1)^2}
\,\,\,\,,\,\,\,\,q^{\nu\beta}_{\bar{1}\bar{2}}
={-\eta^{\nu\beta}\over(y-1)^2}\,\,\,\,,\,\,\,\,
q^{\alpha\beta}_{2\bar{2}}={-\eta^{\alpha\beta}\over4}.
\label{fqy}
\end{eqnarray}
We have dropped terms which will be canceled
when inserted in (\ref{amp1}) using the physical
conditions for closed strings.
Inserting (\ref{fqy}) into (\ref{amp2}) we get
\begin{eqnarray}
A\!\!\! &=&
\!\!\!2i{\mathcal{C}}2^{{\alpha'\over2}
(k_1.D_S.k_1+k_2.D_S.k_2)}
\int^{1}_0 dy(1-y^2)y^{{\alpha\over2}'
k_1.D_S.k_1}(y-1)^{\alpha'k_1.k_2}
(y+1)^{{\alpha'}k_1.D_S.k_2}\cr &&\cr
\!\!\! &\bigg\{&\!\!\! ({\alpha'\over2})^2
\bigg[{a_1\over16y^2}+{a_2\over(y-1)^4}
+{a_3\over(y+1)^4}\bigg]+({\alpha'\over2})^3
\bigg[-{a_4\over16y^2}+{a_5\over4y(y+1)^2}\cr
&&\cr \!\!\! &-& \!\!\!{a_6\over4y(y-1)^2}
-{a_7(y+1)^2\over16y^2(y-1)^2}
-{a_8(y-1)^2\over16y^2(y+1)^2}
+{a_9(y-1)^2\over4y(y+1)^4}
-{a_{10}(y+1)^2\over4y(y-1)^4}\bigg]
\cr &&\cr \!\!\! &+& \!\!\!({\alpha'\over2})^4
\bigg[{a_{11}\over16y^2}
+{a_{12}(y+1)^2\over16y^2(y-1)^2}
+{a_{13}(y-1)^2\over16y^2(y+1)^2}
+{a_{14}(y+1)^4\over16y^2(y-1)^4}
+{a_{15}(y-1)^4\over16y^2(y+1)^4}\bigg]\bigg\}
,\cr &&   \label{amp3}
\end{eqnarray}
where
\begin{equation}
{\mathcal{C}}=C_{D_2}^X g_{c}^2
e^{-\lambda}(2\pi)^{p+1}\delta^{p+1}(k_1
+k_1.D+k_2+k_2.D).
\end{equation}
$D_A$ and $D_S$ are the antisymmetric and
symmetric parts of $D$ matrix. Notice that
because of momentum conservation parallel
to the D-brane we have $k_1.D_A.k_2=0$.
The constants $a_n$'s depend on momentum
and polarization of external states and are
given in the appendix A. We choose the Mandelstam
variables,
\begin{eqnarray}
&&s=-({1\over2}(\eta+D).k_1)^2
=-{1\over2}k_1.D_S.k_1=-{1\over2}k_2.D_S.k_2,
\cr &&\cr && t=-(k_1+k_2)^2=-2k_1.k_2
\,\,\,\,.\label{st}
\end{eqnarray}
Finally making the change of variable
$y={1-\sqrt{x} \over 1+\sqrt{x}}$ we find
\begin{eqnarray}
A\!\!\! &=& \!\!\! {2i{\mathcal{C}}}
({\alpha'\over2})^2 \bigg\{a_1
B(-{\alpha't\over4}+1 ,-\alpha's-1)+a_2
B(-{\alpha't\over4}-1,-\alpha's+1)\cr &&\cr
\!\!\! &+& \!\!\!a_3B(-{\alpha't\over4}+1,-\alpha's+1)
-({\alpha'\over2})\bigg[a_4B(-{\alpha't\over4}
+1,-\alpha's-1)\cr &&\cr \!\!\! &-& \!\!\!
a_5B(-{\alpha't\over4}+1,-\alpha's)
+a_6B(-{\alpha't\over4},-\alpha's)
+a_7B(-{\alpha't\over4},-\alpha's-1)\cr &&\cr
\!\!\! &+& \!\!\!a_8B(-{\alpha't\over4}+2,-\alpha's-1)
-a_9B(-{\alpha't\over4}+2,-\alpha's)\cr &&\cr
\!\!\! &+& \!\!\!a_{10}B(-{\alpha't\over4}-1
,-\alpha's)\bigg]+({\alpha'\over2})^2
\bigg[a_{11}B(-{\alpha't\over4}+1,-\alpha's-1)
\cr &&\cr &+& \!\!\! a_{12}B(-{\alpha't\over4},
-\alpha's-1)+a_{13}B(-{\alpha't\over4}+2,-\alpha's-1)
\cr &&\cr &+& \!\!\! a_{14}B(-{\alpha't\over4}-1
,-\alpha's-1)+a_{15}B(-{\alpha't\over4}+3,
-\alpha's-1)\bigg]\bigg\},\label{accd}
\end{eqnarray}
where $B$'s are beta functions. Note that as a test
of our calculations replacing every polarization of
closed strings by momentum $\epsilon_{\mu\nu}\rightarrow
k_{\mu}k_{\nu}$ gives zero value for the amplitude.
We recover the results of scattering amplitude for
commutative D-branes found in \cite{r15} by turning
the $B$ field off (see Appendix B). As we will be
interested in the low energy effective theory  of
the string, we must analyze the pole structure of
the amplitude (\ref{accd}). There are two types of
singularities in this amplitude that come from the Beta
functions. In fact the amplitude (\ref{accd}) is singular at
\begin{equation}
-{\alpha'\over4}t-1=-n \,\,\,\,,\,\,\,\, -\alpha's-1=-n'
\,\,\,\,,\,\,\,\, n,n'=0,1,\ldots
\end{equation}
which in terms of $t$ and $s$ of (\ref{st}) are
\begin{eqnarray}
&&t=m^{2}_{closed}=-{4\over\alpha'}(-n+1)\cr &&\cr
&&s=m^{2}_{open}=-{1\over\alpha'}(-n'+1),
\end{eqnarray}
exhibiting the mass spectrum of the intermediate
closed string propagators in the t-channel and
the mass spectrum of the open strings in the
s-channel. Note that not all the Beta functions
in (\ref{accd}) have pole for $n=0,1,2,3$ and $n'=0,1$.
Now we may study the low energy limit of
the amplitude i.e. $s$ , $t$ $\rightarrow 0$.
There are then three types of terms to consider\\
$\bf{t-channel:}$
\begin{equation}
A_t={i{\mathcal{C}}\alpha'^2\over
t}\left\{-2a_2s+a_6+a_7+a_{10}
-\alpha'(-a_{10}s+{1\over2}a_{12}+a_{14})
+{\mathcal{O}}(\alpha'^2)\right\},\label{tch}
\end{equation}
$\bf{s-channel:}$
\begin{eqnarray}\label{sch}
A_s\!\!\! &=& \!\!\!{i{\mathcal{C}}\alpha'^2\over
-4s}\bigg\{{1\over2}a_1t+a_5-a_6-a_7
+a_8+a_9-a_{10}-({\alpha'\over4})\bigg[(a_4
+a_7+a_8)t\cr &&\cr\!\!\! &-& \!\!\!
2a_{12}+2a_{13}-4a_{14}+4a_{15}\bigg]
+{\mathcal{O}}(\alpha'^2)\bigg\},
\end{eqnarray}
$\bf{contact-terms:}$
\begin{eqnarray}\label{cch}
A_c\!\!\! &=&
\!\!\!{i{\mathcal{C}}\alpha'^2\over2}
\bigg\{-(a_1+a_2-a_3)+({\alpha'\over2})
\bigg[a_4+a_7+a_8-a_9+a_{10}\cr &&\cr
\!\!\! &-& \!\!\! (-2s-{1\over2}t)(a_1
+a_2+a_3)\bigg]+{\mathcal{O}}(\alpha'^2)\bigg\}.
\end{eqnarray}
\section{Low energy effective action}
In this section we will derive the low energy
action for the above amplitude to the first two
leading orders of $\alpha'$. For scattering of two
gravitons we use the physical condition for them as,
\begin{equation} \epsilon_{\mu\nu}=\epsilon_{\nu\mu}
\,\,\,\,,\,\,\,\,{\epsilon^{\mu}}_{\mu}=0.
\end{equation}

In the t-channel we have two gravitons which
interact in the bulk space and produce another
graviton or dilaton field which in turn is
absorbed by the D-brane. It is easily seen that the
massless antisymmetric field does not contribute
to the amplitude. The t-channel amplitude to be
reproduced by inserting for the coefficients $a_{i}$
of appendix A,
\begin{eqnarray}
A_t\!\!\! &=& \!\!\!{i{\mathcal{C}}\alpha'^2\over
t}\bigg\{(k_1.D_S.k_1)Tr(\epsilon_1.\epsilon_2)
+Tr(\epsilon_1.D_S)(k_1.\epsilon_2.k_1)
+Tr(\epsilon_2.D_S)(k_2.\epsilon_1.k_2)
\cr &&\cr
\!\!\! &-&2k_2.\epsilon_1.D_S.\epsilon_2.k_1
- 2k_1.\epsilon_2.\epsilon_1.k_2
-2k_1.D_S.\epsilon_1.\epsilon_2.k_1
-2k_2.D_S.\epsilon_2.\epsilon_1.k_2
 \cr &&\cr &
-&\alpha'\bigg[(k_1.\epsilon_2.\epsilon_1
.k_2)(k_1.D_S.k_1)+(k_1.\epsilon_2
.k_1)(k_2.\epsilon_1.k_2)+(k_1.\epsilon_2
.k_1)(k_2.\epsilon_1.D_S.k_2) \cr &&\cr
\!\!\! &+&\!\!\!(k_2.\epsilon_1.k_2)
(k_1.\epsilon_2.D_S.k_1)\bigg]\bigg\}=
A_{t}^0+A_{t}^1, \label{lowt}
\end{eqnarray}
where $A_{t}^0 (A_{t}^1)$ contains
terms in the curly bracket which are zeroth
(first) order at $\alpha'$. It is not difficult
to see that the action which describes this
amplitude to order of $\alpha'^0$ is nothing
but the DBI and Einstein-Hilbert action.
The DBI action in the Einstein frame is
,\footnote{For simplicity in writing we
have dropped the coefficient of $f$ i.e.
$2\pi\alpha'f\rightarrow f.$}
\begin{equation}
{\mathcal{S}}_{D-brane}^0=-T_p\int d^{p+1}x
e^{-\Phi}\sqrt{-det(e^{-\gamma \Phi
}g+{\mathcal{B}}+f)_{ab}}\,\,\,\,\,,\label{dbi}
\end{equation}
where $g$ is the induced metric on the D-brane and
${\mathcal{B}}_{ab}=B_{ab}-2\kappa b_{ab}$
is the pull back of the antisymmetric field along
the D-brane with $B$ constant and $f_{ab}$ is the
gauge field strength on the D-brane and
$\gamma=-{4 \over d-2}$. Expanding $g_{ab}$
around the Minkowski metric, $g_{ab}=\eta_{ab}
+2\kappa h_{ab}$ we get for the action to the
first order of $h$,\footnote{Note that we have
expanded the action (\ref{dbi}) around the
background $\eta_{ab}+B_{ab}$ using $\sqrt{det(M_0
+M)}=\sqrt{detM_0}(1+{1\over2}Tr(M_{0}^{-1}M)+...).$}
\begin{equation}
{\mathcal{S}}_{D-brane}^0=-\kappa T_p c\int d^{p+1}x
h_{ab}V^{ab} ,\label{dbiv}
\end{equation}
where
\begin{equation}
V^{ab}={1\over2}(\eta+D)^{ba}\,\,\,\,,\,\,\,\,
c=\sqrt{-det(\eta+B)_{ab}}\,\,\,\,\,.
\end{equation}
Equation (\ref{dbiv}) exhibits a source term for gravity on the
D-brane,
\begin{equation}
({S}_h)^{ab}={-{1\over2}T_p\kappa c}(\eta^{ab}+D_S^{ab}).
\end{equation}
The Einstein-Hilbert action in the bulk is
\begin{equation}
S^{(0)}_{bulk}={1\over 2\kappa^2}\int d^{26}x \sqrt{-G}R.
\end{equation}
Using 26-dimensional propagator $(G^{hh})
_{\mu\nu\lambda\rho}$ and the three point
interaction vertex $(V_{h\epsilon_1
\epsilon_2})^{\lambda\rho}$ for gravitons
coming from this action \cite{r6}, and the
above source term, we find
\begin{equation}
i(S_h)^{\mu\nu}(G^{hh})_{\mu\nu\lambda\rho}(
V_{h\epsilon_1\epsilon_2})^{\lambda\rho}
=A_{t}^0+C_{t}^0,
\end{equation}
where $A_{t}^0$ is exactly the zeroth order
term of the amplitude (\ref{lowt}), and
$C_{t}^0$ is a contact term with no poles,
\begin{equation}
C_{t}^0={-i c T_p \kappa^2}\left(
Tr(\epsilon_1.\epsilon_2)+Tr(\epsilon_1.
D_S.\epsilon_2)\right).\label{ct0}
\end{equation}
Here the tension of the D-branes, $T_p$
from (\ref{dbi}) is written in terms of
the coefficients of (\ref{lowt}) i.e.
\begin{equation}
T_p={{\mathcal{C}}\alpha'^2\over c\kappa^2}.
\end{equation}
We note that to this order of $\alpha'$ the
above results are exactly the same as in the
superstring theory \cite{r6}. To get the next
order terms in the string amplitude (\ref{lowt}), we
include the next order of $\alpha'$ gravitational
action in the bulk, \cite{r3,r4}
\begin{equation}
S^{(1)}_{bulk}={\alpha' \over 8{\kappa}^2}\int d^{26}x
e^{\gamma\Phi}\sqrt{-G}(R^{\mu\nu\kappa\lambda}
R_{\mu\nu\kappa\lambda}-4R^{\mu\nu}R_{\mu\nu}
+R^2).\label{bulk1}
\end{equation}
There is a new three point interaction for gravitons
$V'_{h\epsilon_1\epsilon_2}$ in this Lagrangian.
In addition it contains the interaction between
one dilaton and two gravitons $V_{\Phi\epsilon_1
\epsilon_2}$. So there are two source terms,
gravitational source $S_h$ and dilatonic source
$S_{\Phi}$ and we find
\begin{equation}
iS_{h}G^{hh}V'_{h\epsilon_{1}\epsilon_{2}}
+iS_{\Phi}G^{\Phi\Phi}V_{\Phi\epsilon_{1}\epsilon_{2}}
=A_{t}^{1}+C_{t}^{1},
\end{equation}
where $A_{t}^1$ is the first order term of the amplitude
(\ref{lowt}) and
\begin{eqnarray}
C_{t}^1\!\!\! &=& \!\!\!{i c T_p \kappa^2\alpha'\over
2}\bigg\{-k_1.D_S.k_1Tr(\epsilon_1.\epsilon_2)
+k_1.k_2Tr(\epsilon_1.D_S.\epsilon_2)
-k_1.\epsilon_2.\epsilon_1.D_S.k_2\cr &&\cr
\!\!\! &+& \!\!\!k_2.D_S.\epsilon_2.\epsilon_1.k_2
-k_1.\epsilon_2.D_S.\epsilon_1.k_2)\bigg\},\label{ct1}
\end{eqnarray}
is a contact term. This calculation exhibits
similar gravitational source term as in the
order of $\alpha'^0$.

Next we consider the s-channel amplitude up
to the first two powers of $\alpha'$.
We may write this amplitude, (\ref{sch}), as
\begin{eqnarray}
A_s\!\!\! &=& \!\!\!{-ic\kappa^2T_p\over 4s}
\bigg\{Tr(\epsilon_1.D_S)(k_1.
\epsilon_2.D_S.k_2-k_1.D_S.\epsilon_2.
D_S.k_2-k_1.D_A.\epsilon_2.D_A.k_2)
\cr &&\cr \!\!\! &+& \!\!\! k_2.D_A.\epsilon_2.
D_S.\epsilon_1.D_A.k_1+k_2.D_S.\epsilon_2.
D_S.\epsilon_1.D_S.k_1+2k_2.D_A.\epsilon_2.
D_A.\epsilon_1.D_S.k_1\cr &&\cr \!\!\! &-& \!\!\!
k_2.D_S.\epsilon_2.\epsilon_1.D_S.k_1+
k_2.D_A.\epsilon_2.\epsilon_1.D_A.k_1
-{1\over2}k_1.k_2Tr(\epsilon_1.D_S)
Tr(\epsilon_2.D_S)\cr &&\cr \!\!\! &+& \!\!\!
({\alpha'\over2})\bigg[k_1.k_2 Tr(\epsilon_1.
D_S)(k_2.D.\epsilon_2.D.k_2)+(k_2.
\epsilon_1.D_S.k_1)(k_1.\epsilon_2.
D_S.k_2)\cr &&\cr \!\!\! &-& \!\!\!
(k_1.D_S.\epsilon_1.D_S.k_2)
(k_2.D_S.\epsilon_2.D_S.k_1)
+2(k_1.D_A.\epsilon_2.D_A.k_1)
(k_1.D_S.\epsilon_1.D_S.k_2)\cr &&\cr
\!\!\! &-& \!\!\! (k_1.D_A.\epsilon_2.D_A.k_1)
(k_2.D_A.\epsilon_1.D_A.k_2)\bigg]\bigg\}+
(1\leftrightarrow2)=A_{s}^0+A_{s}^1. \label{lows}
\end{eqnarray}
In this channel we expect the two gravitons
scatter from the D-brane and exchange a gauge
or scalar field which propagates on the D-brane.
These propagating fields are in fact the low energy
limits of the massless open strings on the D-brane,
scalars $(X^i)$ and gauge bosons $(a_a)$.
We would like to write down a Lagrangian for these
fields on the D-brane which produces the vertices
for interactions between gravitons and these fields
and finally gives the scattering amplitude of the
string theory. For zeroth order of $\alpha'$ we
again take the DBI action on D-brane (\ref{dbi})
with gauge field strength defined as
$f_{ab}=\partial_a a_b-\partial_b a_a$
then expand the induced metric
$g_{ab}=\partial_a X^{\mu}\partial_b X^{\nu}g_{\mu\nu}$
on the D-brane around the flat space and choose
the static gauge $(X^a=x^a)$ to get
\begin{equation}
g_{ab}=\eta_{ab}+2\kappa
h_{ab}+2\kappa(h_{ia}\partial_b X^i+h_{bi}
\partial_a X^i)+\partial_a X^i
\partial_b X_i+2\kappa h_{ij}\partial_a
X^i\partial_b X^j.\label{indg}
\end{equation}
Redefining the gauge and scalar fields by $A_a=
2\pi\alpha'\sqrt{T_p}a_a$ and
$ \lambda^i=\sqrt{T_p}X^i$ and expanding
DBI action around the background field
$\eta_{ab}+B_{ab}$, for the zeroth order of
$\alpha'$ we find the following vertices,
\begin{eqnarray}
&&(V_{\epsilon_1A})^{a}=\sqrt{T_p}\kappa
c\left(Tr(\epsilon_1.V_S)k_1.V_{A}^a
-2k_1.V_S.\epsilon_1.V_{A}^a
-2k_1.V_A.\epsilon_1.V_{S}^a\right),\cr &&\cr
&&(V_{\epsilon_1\lambda})_i=\sqrt{T_p}\kappa
c\left(Tr(\epsilon_1.V_S)k_{1i}-2k_1.V_S
.\epsilon_{1i}\right),
\end{eqnarray}
and propagators,
\begin{equation}
(G^{\lambda\lambda})^{ij}={i\over c s}\eta^{ij}
\,\,\,\,,\,\,\,\, (G^{AA})_{ab}={i\over c s}
(V_{S}^{-1})_{ab}\,\,\,\,,\,\,\,\,
\end{equation}
where $V_A$ and $V_S$ are the antisymmetric
and symmetric parts of the  matrix $V$.
Note that in deriving the above vertices we
have used the fact that the closed string
fields in the DBI action appear as functionals
of the transverse scalar fields $X^i$
which should be Taylor expanded. Knowing
these vertices and propagators we can compute
the amplitude from the effective field theory
on D-brane and show that to order of $\alpha'^0$,
\begin{equation}
V_{\epsilon_1\lambda}G^{\lambda\lambda}V_{\lambda
\epsilon_2}+V_{\epsilon_1A}G^{AA}V_{A\epsilon_2}
=A_{s}^0+C_{s}^0,
\end{equation}
where $A_{s}^0$ is the zeroth order term in
(\ref{lows}) and $C_{s}^0$ is a contact term,
\begin{equation}
C_{s}^0={-i\kappa^2T_pc\over4}
Tr(\epsilon_1.D_S)Tr(\epsilon_2.D_S).\label{cs0}
\end{equation}
We have used the following useful relations
in the derivation
\begin{equation}
(V_AV_S^{-1}V_A)^{ab}=-{1\over2}(\eta-D_S)^{ab}
\,\,\,\,,\,\,\,\,\eta^{ij}={1\over2}(\eta-D_S)^{ij}.
\end{equation}
It turns out that in order to describe the
amplitude to the first order in $\alpha'$
we need to add new terms to the DBI action as follow
\pagebreak
\begin{eqnarray}
{\mathcal{S}}^1\!\!\! &=& \!\!\!
{-\alpha'T_p\over2}\int d^{p+1}x \bigg
\{ \sqrt{-det(\eta+B+f)}R_{abcd}({1\over \eta+B+f})^{ad}
({1\over g+B+f})^{bc}\cr &&\cr
\!\!\! &-& \!\!\! \sqrt{-det(g+B+f)}
({\Omega^i}_{ac}{\Omega_i}_{bd}
-{\Omega^i}_{ad}{\Omega_i}_{bc})
({1\over g+B+f})^{ad} ({1\over
g+B+f})^{bc}\bigg\}, \cr &&\label{act}
\end{eqnarray}
where ${R}_{abcd}=g_{ae}{R^e}_{bcd}$ and ${R^e}_{bcd}$
is Riemann tensor constructed out of induced
metric on the D-brane and $\Omega$'s are second
fundamental forms,
\begin{eqnarray}
{R^a}_{bcd}\!\!\! &=&
\!\!\!{\Gamma^a}_{bd,c}-{\Gamma^a}_{bc,d}+{\Gamma^e}_{bd}
{\Gamma^a}_{ce}-{\Gamma^e}_{bc}{\Gamma^a}_{de}\,\,\,\,,
\cr &&\cr \Omega^{i}_{ab}\!\!\! &=&
\!\!\!\kappa(-\partial^i h_{ab}+\partial_a {h^i}_b
+\partial_b{h^i}_a)+\partial_a
\partial_b X^i\,\,\,\,,
\end{eqnarray}
which  $a,b,c,d,e=0, \ldots,p $ and $ i,j=p+1,\ldots d$.
The $\Gamma$'s are the Christoffel symbols constructed
out of the induced metric $g$. This action contains
every vertex that we need for producing the next order
of $\alpha'$ in (\ref{lows}), in addition to certain
contact terms which we will describe later. To reproduce
the s-channel amplitude of string theory from such an
action we need to find the graviton-scalar and graviton-gauge
field vertices. We find the first one from the second
fundamental form linear in terms of $X$ and $h$,
\footnote{We must Taylor expand the linear terms in
$h$ of the above action, but because of
momentum conservation relation these terms do not
contribute to the vertex.}and the second one from
the expansion the above Lagrangian linear in terms
of $f$ and $h$. By summing these vertices corresponding
to the above action with the previous terms which we
found from DBI action to zeroth order of $\alpha'$,
we find that
\begin{eqnarray}
(V_{\epsilon_1A})^a\!\!\! &=& \!\!\!{\kappa c
\sqrt{T_p}\over4}\bigg\{2Tr(\epsilon_1.
D_S)k_1.V_{A}^a-4k_1.D_S.\epsilon_1.V_{A}^a
+4k_1.D_A.\epsilon_1.V_{S}^a\cr &&\cr
\!\!\! &-& \!\!\! \alpha'\bigg[(k_1.D_S.
\epsilon_1.D_S.k_1+k_1.D_A.\epsilon_1.
D_A.k_1)k_1.V_{A}^a-k_1.D_S.k_1
Tr(\epsilon_1.D_S)k_1.V_{A}^a\bigg]\bigg\}\cr &&\cr
(V_{\epsilon_1\lambda})_i\!\!\! &=& \!\!\!{\kappa c
\sqrt{T_p}\over4}
\bigg\{2Tr(\epsilon_1.D_S)k_{1i}-4k_1.D_S.
\epsilon_{1i}\cr &&\cr\!\!\! &-& \!\!\! \alpha'
\bigg[(k_1.D_S.\epsilon_1.D_S.k_1
+k_1.D_A.\epsilon_1.D_A.k_1)k_{1i}
-k_1.D_S.k_1Tr(\epsilon_1.D_S)
k_{1i}\bigg]\bigg\}.\label{ver}
\end{eqnarray}
To check that these vertices are in fact correct
we have calculated one closed and one open string
scattering from D-brane and we have found similar
vertices (see Appendix C). Now using these vertices
we can find the first order of $\alpha'$ terms in
s-channel from string theory. We have used the previous
propagators for gauge and scalar fields, as (\ref{act})
does not introduce any corrections to the propagators.
We then find
\begin{equation}
V_{\epsilon_1\lambda}G^{\lambda\lambda}V_{\lambda
\epsilon_2}+V_{\epsilon_1A}G^{AA}V_{A\epsilon_2}=
A_{s}^0+C_{s}^0+A_{s}^1+C_{s}^1,
\end{equation}
in which $A_{s}^0(A_{s}^1)$ is exactly the
zeroth (first) order of $\alpha'$ terms of
(\ref{lows}), $C_{s}^0$ is the zeroth order term in
(\ref{cs0}) and $C_{s}^1$ is the first order in
$\alpha'$ terms with no poles and can be written as
\begin{eqnarray}
C_{s}^1\!\!\! &=& \!\!\!{ic\kappa^2T_p
\alpha'\over4}\bigg\{{1\over2}Tr(\epsilon_1.D_S)
Tr(\epsilon_2.D_S)(k_1.D_S.k_2-k_1.k_2)
+\cr &&\cr\!\!\! &+&\!\!\!Tr(\epsilon_1.D_S)
\bigg[2k_2.D_S.\epsilon_2.k_1
+{3\over2}k_2.D.\epsilon_2.D.k_2\bigg]\bigg\}
+(1\leftrightarrow2).\label{cs1}
\end{eqnarray}
It remains to verify that all the above contact
terms (\ref{ct0}), (\ref{cs0}) and contact terms
from (\ref{dbi}) at $\alpha'^0$ and (\ref{ct1})
, (\ref{cs1}) and contact terms from (\ref{act})
at $\alpha'$ add up to the contact terms of the
string amplitude (\ref{cch}). We can write
(\ref{cch}) as
\begin{eqnarray}
A_{c}\!\!\! &=& \!\!\!{ic\kappa^2T_p\over4}
\bigg\{-Tr(\epsilon_1.D)Tr(\epsilon_2.D)
-Tr(\epsilon_1.\epsilon_2)
+Tr(D.\epsilon_1.D.\epsilon_2)\cr &&\cr
\!\!\! &+&
\!\!\!\alpha'\bigg[Tr(\epsilon_1.D)(k_2.D.
\epsilon_2.D.k_2)-k_1.D.\epsilon_2.D.
\epsilon_1.D.k_2+k_1.\epsilon_2.
\epsilon_1.k_2\cr &&\cr \!\!\! &+& \!\!\!
{1\over2}(k_1.D.k_2) (Tr(\epsilon_1.D)
Tr(\epsilon_2.D)+Tr(\epsilon_1.\epsilon_2)
+Tr(D.\epsilon_1.D.\epsilon_2)\bigg]\bigg\}\cr
&&\cr &+&(1\longleftrightarrow2).\label{lowc}
\end{eqnarray}
If we expand DBI action (\ref{dbi}), we find
graviton-graviton contact terms from this expansion,
\begin{equation}
A_{hh}^0=-2i\kappa^2cT_{p}\bigg\{{1\over2}
Tr(\epsilon_1.V)Tr(\epsilon_2.V)
-Tr(\epsilon_1.V.\epsilon_2.V)\bigg\}.
\end{equation}
We can show that by adding (\ref{ct0}) and
(\ref{cs0}) to the above equation
\begin{equation}
A_{hh}^0+C_{t}^0+C_{s}^0={-i\kappa^2
cT_p\over2}\bigg\{Tr(\epsilon_1.D)
Tr(\epsilon_2.D)+Tr(\epsilon_1.\epsilon_2)
-Tr(\epsilon_1.D.\epsilon_2.D)\bigg\},
\end{equation}
we find exactly the $\alpha'^0$ term in (\ref{lowc}).
Again as before if we expand
Lagrangian in (\ref{act}), we find for
graviton-graviton contact terms,
\begin{eqnarray}
A_{hh}^1\!\!\! &=& \!\!\!-2i\alpha'\kappa^2cT_p
\bigg\{-k_1.V.\epsilon_2.\epsilon_1.V.k_2
-k_1.V.\epsilon_2.V.\epsilon_1.k_2
-k_1.\epsilon_2.V.\epsilon_1.V.k_2\cr &&\cr
&+&k_1.V.\epsilon_2.V.\epsilon_1.V.k_2
-{1\over4}k_1.k_2Tr(\epsilon_1.V)Tr(\epsilon_2.V)
+{1\over2}k_1.V.k_2Tr(\epsilon_1.V.
\epsilon_2)\cr &&\cr &+&{1\over4}k_1.k_2
Tr(V.\epsilon_1.V.\epsilon_2)+{1\over2}k_1
.V.k_1Tr(V.\epsilon_1.V.\epsilon_2)\cr &&\cr
\!\!\! &+& \!\!\!\left(k_1.V.\epsilon_1.k_2
+{1\over2}k_1.V.\epsilon_1.V.k_1\right)
Tr(\epsilon_2.V)\bigg\}+(1\leftrightarrow2),
\end{eqnarray}
we can show that by adding (\ref{ct1}) and
(\ref{cs1}) to above equation,
\pagebreak
\begin{eqnarray}
A_{hh}^1+C_{t}^1+C_{s}^1  \!\!\! &=& \!\!\!
{i\alpha'\kappa^2cT_p\over4}
\bigg\{Tr(\epsilon_1.D)k_2.D.\epsilon_2.D.k_2
-k_1.D.\epsilon_2.D.\epsilon_1.D.k_2+
k_1.\epsilon_2.\epsilon_1.k_2\cr &&\cr
\!\!\! &+& \!\!\!{1\over2}(k_1.D.k_2)
\bigg[Tr(\epsilon_1.D)Tr(\epsilon_2.D)
+Tr(\epsilon_1.\epsilon_2)+Tr(D.\epsilon_1.
D.\epsilon_2)\bigg] \bigg\}\cr &&\cr
\!\!\! &+& \!\!\!(1\longleftrightarrow2),
\end{eqnarray}
we recover exactly the $\alpha'$ term in (\ref{lowc}).
It is noticeable that in evaluating the t-channel
poles in field theory we use the linear coupling
of graviton to the D-brane using the DBI action
(\ref{dbi}) which is of zero order of $\alpha'$
whereas linear coupling of graviton to D-brane
coming from action (\ref{act}) is of first order
of $\alpha'$ however using conservation of momentum
this coupling has no effect in the t-channel.

It should be noted that when we turn off the $B$
field on the D-brane we find the Einstein-Hilbert
action as we expected \cite{r15}. To see this,
note that as we turn off the $B$ field on
the D-brane we see, $V^{ab}=\eta^{ab}$ where upon
(\ref{act}) changes up to some total derivatives to
\begin{eqnarray}
{\mathcal{S}}^1_{D-brane}\!\!\! &=& \!\!\!
-{\alpha'T_p\over2}\int d^{p+1}x
\sqrt{-detg_{ab}}(R+{{\Omega^i}_a}^a
{{\Omega_i}_b}^b-{\Omega^i}_{ab}
{\Omega_i}^{ab}),\cr &&\cr
R\!\!\! &=& \!\!\!R_{abcd}g^{ac}g^{bd}, \label{actb0}
\end{eqnarray}
which  $a,b,c,d=0, \ldots,p $ and $ i,j=p+1,\ldots d$.
Where we have ignored gauge fields because they do not
have any contraction with gravitons because of the
antisymmetry of $f_{ab}$. In deriving the
above equation we have used the following relation at
${\mathcal{O}}(h^2)$,
\begin{equation}
\sqrt{-det\eta}R_{abcd}\eta^{ad}g^{bc}
=\sqrt{-detg}R_{abcd}g^{ac}g^{bd}+(T.D),
\end{equation}
which by momentum conservation relation,
total derivative terms
have no effect in scattering amplitude.
\section{Discussion}
In this work we have found the corrections to
the DBI action to order of $\alpha'$,
(\ref{act}), for the description of the graviton-
graviton-D-brane scattering in bosonic string
theory in the presence of a constant $B$ field.

We have also calculated the corrections to
the DBI action for dilaton and antisymmetric
field scattering with a D-brane.
The results are not qualitatively
different and will be reported separately.

It is known that D-branes in the presence of
background $B$ field become noncommutative
and the DBI action can be written in terms of
commutative gauge fields or in terms of the
noncommutative open string fields involve the
$\star$ product and a Wilson line operator.
We write the action (\ref{act}) in terms
of commutative fields.

One may write the result in eq(\ref{act})
in the following form,
\pagebreak
\begin{eqnarray}
{\mathcal{S}}^1 \!\!\! &=& \!\!\! {-\alpha'
T_p\over2}\int d^{p+1}x \sqrt{-det(\eta+B+f)}
({1\over \eta+B+f})^{ad} ({1\over g+B+f})^{bc}
\cr &&\cr \!\!\! & \times & \!\!\!
\left\{R_{abcd}-{\Omega^i}_{ac}{\Omega_i}_{bd}
+{\Omega^i}_{ad}{\Omega_i}_{bc} \right\},
\label{dact}
\end{eqnarray}
which is reminiscent of the corresponding result
for the case $B=0$, eq (\ref{actb0}). This action
as well as the action in (\ref{act}) are not in
covariant form. One may try to write it in a
covariant form;
\begin{eqnarray}
{\mathcal{S}}^1_{cov} \!\!\! &=& \!\!\!
{-\alpha'T_p\over2}\int d^{p+1}x \sqrt{-det(g+B+f)}
({1\over g+B+f})^{ad} ({1\over g+B+f})^{bc} \cr &&\cr
\!\!\! & \times & \!\!\!
\bigg\{R_{abcd}-{\Omega^i}_{ac}{\Omega_i}_{bd}
+{\Omega^i}_{ad}{\Omega_i}_{bc}\bigg\}; \label{covact}
\end{eqnarray}
but, this will give rise to new contact terms which are not present
in the string theory amplitude, eq. (\ref{lowc}).
However, on general grounds one expects a covariant form for the action.
It is conceivable that addition of terms involving covariant derivatives of
the gauge field f contracted with the metric g or g+B may conspire to cancel
these unwanted terms and reproduce the string amplitude.

In view of the large variety of such possible terms and the complication
of these calculations, we have not been able to find a simple covariant
variation of the action (\ref{dact}). The same difficulties arise for
application of T duality on the action which would be a check of our result.

\section*{Acknowledgments}

We would like to thank M. Alishahiha and
S. F. Hassan for useful conversations.
F.A acknowledges discussion with L. Alvarez-Gaume.

\pagebreak

{\large\bf{Appendix A}} \\
{\bf Coefficients for Noncommutative D-brane}

Here we presented the coefficient for scattering
of two closed string from D-brane in the presence
of $B$ field.
\begin{eqnarray}
a_1\!\!\! &=& \!\!\!Tr(\epsilon_1.D)
Tr(\epsilon_2.D)\cr &&\cr a_2\!\!\! &=& \!\!\!
Tr(\epsilon_{1}^T.\epsilon_2)\cr &&\cr
a_3\!\!\! &=& \!\!\!Tr(D.\epsilon_1.D.
\epsilon_2)\cr &&\cr a_4\!\!\! &=& \!\!\!
Tr(\epsilon_1.D)(k_1.\epsilon_2.D.k_1
+k_1.D.\epsilon_2.k_1)
+(1\longleftrightarrow2)\cr &&\cr
a_5\!\!\! &=& \!\!\!k_1.\epsilon_2.D.
\epsilon_1D.k_2
+k_2.\epsilon_1.D.\epsilon_2.D.k_1
+k_1.D.\epsilon_2.D.\epsilon_1.k_2
+k_2.D.\epsilon_1.D.\epsilon_2.k_1
\cr &&\cr
\!\!\! \!\!\! &-&  \!\!\!k_2.D^T.\epsilon_{1}
^T.\epsilon_2.D.k_1-k_2.D.\epsilon_1.
\epsilon_{2}^T.D^T.k_1\cr &&\cr a_6
\!\!\! &=& \!\!\!k_2.D^T.\epsilon_{1}^T.
\epsilon_2.k_1+k_2.\epsilon_{1}^T.
\epsilon_2.D.k_1-k_1.\epsilon_2.D.
\epsilon_1.k_2-k_2.\epsilon_1.D.
\epsilon_2.k_1\cr &&\cr
\!\!\! &+& \!\!\!k_2.\epsilon_1.
\epsilon_{2}^T.D^T.k_1+k_2.D.\epsilon_1.
\epsilon_{2}^T.k_1\cr &&\cr
a_7\!\!\! &=& \!\!\!Tr(\epsilon_1.D)
(k_1.\epsilon_2.k_1)+
(1\longleftrightarrow2)\cr &&\cr
a_8\!\!\! &=& \!\!\!Tr(\epsilon_1.D)
(k_1.D.\epsilon_2.D.k_1)
+(1\longleftrightarrow2)\cr &&\cr
a_9\!\!\! &=& \!\!\!k_1.D.\epsilon_2.
D.\epsilon_1.D.k_2+(1\longleftrightarrow2)
\cr &&\cr a_{10}\!\!\! &=& \!\!\!k_2.
\epsilon_{1}^T.\epsilon_2.k_1
+k_2.\epsilon_1.\epsilon_{2}^T.k_1\cr &&\cr
a_{11}\!\!\! &=& \!\!\!(k_1.\epsilon_2.D.k_1+
k_1.D.\epsilon_2.k_1)(k_2.\epsilon_1.D.k_2
+k_2.D.\epsilon_1.k_2)\cr &&\cr
\!\!\! &+& \!\!\!(k_1.\epsilon_2.k_1)(k_2.D.
\epsilon_1.D.k_2)+(k_1.D.\epsilon_2.D.k_1)
(k_2.\epsilon_1.k_2)\cr &&\cr a_{12}\!\!\! &=&
\!\!\!(k_1.\epsilon_2.k_1)(k_2.\epsilon_1
.D.k_2+k_2.D.\epsilon_1.k_2)
+(1\longleftrightarrow2)\cr &&\cr a_{13}\!\!\!&=&
\!\!\!(k_1.D.\epsilon_2.D.k_1)
(k_2.\epsilon_1.D.k_2+k_2.D.\epsilon_1.k_2)
+(1\longleftrightarrow2)\cr &&\cr
a_{14}\!\!\! &=& \!\!\!(k_1.\epsilon_2.k_1)
(k_2.\epsilon_1.k_2)\cr &&\cr
a_{15}\!\!\! &=&\!\!\!(k_1.D.\epsilon_2.D.k_1)
(k_2.D.\epsilon_1.D.k_2).
\end{eqnarray}\\
\pagebreak

{\large\bf{Appendix B}} \\
{\bf Coefficients for commutative D-brane}

When we turn the $B$ field off we must only change
$D_{S}^{ab}=D_{0}^{ab}=\eta^{ab}$ and
$D_{S}^{ij}=D_{0}^{ij}=-\delta^{ij} $ and $D_A=0$
again we find the same amplitude as (\ref{accd})
by including these changes and
replacing $a_n$ coefficients with $b_n$, where
\begin{eqnarray}
b_1\!\!\! &=& \!\!\!Tr(\epsilon_1.D_0)
Tr(\epsilon_2.D_0)\cr &&\cr
b_2\!\!\! &=& \!\!\!Tr(\epsilon_{1}^T.
\epsilon_2)\cr &&\cr b_3\!\!\! &=& \!\!\!
Tr(D_0.\epsilon_1.D_0.\epsilon_2)\cr &&\cr
b_4\!\!\! &=& \!\!\!Tr(\epsilon_1.D_0)
(k_1.\epsilon_2.D_0.k_1+k_1.D_0.
\epsilon_2.k_1)+(1\longleftrightarrow2)\cr &&\cr
b_5\!\!\! &=& \!\!\!k_1.\epsilon_2.D_0.
\epsilon_1.D_0.k_2+k_2.\epsilon_1.D_0.
\epsilon_2.D_0.k_1+k_1.D_0.\epsilon_2.
D_0.\epsilon_1.k_2+k_2.D_0.\epsilon_1.
D_0.\epsilon_2.k_1\cr &&\cr\!\!\! \!\!\! &-&
\!\!\!k_2.D_0.\epsilon_{1}^T.\epsilon_2
.D_0.k_1-k_2.D_0.\epsilon_1.\epsilon_{2}
^T.D_0.k_1\cr &&\cr b_6\!\!\! &=& \!\!\!
k_2.D_0.\epsilon_{1}^T.\epsilon_2.k_1
+k_2.\epsilon_{1}^T.\epsilon_2.D_0.k_1
-k_1.\epsilon_2.D_0.\epsilon_1.k_2
-k_2.\epsilon_1.D_0.\epsilon_2.k_1\cr &&\cr
\!\!\! &+& \!\!\!k_2.\epsilon_1.\epsilon_{2}
^T.D_0.k_1+k_2.D_0.\epsilon_1.\epsilon_{2}
^T.k_1 \cr &&\cr b_7\!\!\! &=&
\!\!\!Tr(\epsilon_1.D_0)(k_1.\epsilon_2.k_1)
+(1\longleftrightarrow2)\cr &&\cr
b_8\!\!\! &=& \!\!\!Tr(\epsilon_1.D_0)
(k_1.D_0.\epsilon_2.D_0.k_1)+
(1\longleftrightarrow2)\cr &&\cr b_9\!\!\! &=&
\!\!\!k_1.D_0.\epsilon_2.D_0.\epsilon_1
.D_0.k_2 +(1\longleftrightarrow2)\cr &&\cr
b_{10}\!\!\! &=& \!\!\!k_2.\epsilon_{1}^T.
\epsilon_2.k_1+k_2.\epsilon_1.\epsilon_{2}
^T.k_1\cr &&\cr b_{11}\!\!\! &=& \!\!\!
(k_1.\epsilon_2.D_0.k_1
+k_1.D_0.\epsilon_2.k_1)
(k_2.\epsilon_1.D_0.k_2+k_2.D_0.
\epsilon_1.k_2)\cr &&\cr\!\!\! &+& \!\!\!
(k_1.\epsilon_2.k_1)(k_2.D_0.\epsilon_1
.D_0.k_2)+(k_1.D_0.\epsilon_2.D_0.k_1)
(k_2.\epsilon_1.k_2)\cr &&\cr b_{12}\!\!\! &=&
\!\!\!(k_1.\epsilon_2.k_1)(k_2.
\epsilon_1.D_0.k_2+k_2.D_0.\epsilon_1.
k_2)+(1\longleftrightarrow2) \cr &&\cr
b_{13}\!\!\! &=& \!\!\!(k_1.D_0.
\epsilon_2.D_0.k_1)(k_2.\epsilon_1.
D_0.k_2 +k_2.D_0.\epsilon_1.k_2)
+(1\longleftrightarrow2)\cr &&\cr
b_{14}\!\!\! &=& \!\!\!(k_1.\epsilon_2.k_1)
(k_2.\epsilon_1.k_2)\cr &&\cr
b_{15}\!\!\! &=& \!\!\!(k_1.D_0.\epsilon_2
.D_0.k_1)(k_2.D_0.\epsilon_1.D_0.k_2).
\end{eqnarray} \\
\pagebreak

{\large\bf{Appendix C}} \\
{\bf One open and one closed bosonic string scattering}

In this appendix we want to calculate scattering
of one massless open string with one massless
closed  graviton string. In the bosonic case
vertices for open and closed string are
\begin{eqnarray}
V_c\!\!\! &=&
\!\!\!\epsilon_{\mu\lambda}{D^\lambda}_\nu:\partial
X^\mu(z_1)e^{ik_1.X(z_1)}::\bar{\partial}
X^\nu(\bar{z}_1)e^{ik_1.D.X(\bar{z}_1)}:,\cr &&\cr
V_o\!\!\! &=&
\!\!\!{\xi}_\alpha{(V^T)^{\alpha}}_{\rho}:\partial
X^\rho(y) e^{2ik_2.V^T. X(y)}:,
\end{eqnarray}
where again we have used doubling trick for writing them.
For momentum conservation relation we have
\begin{equation}
k_{1}^\mu+k_1.D^\mu+2k_2.V^{T\mu}=0,
\end{equation}
and physical conditions for momentum and polarizations are
\begin{eqnarray}
k_1.\epsilon^\mu\!\!\! &=& \!\!\!0
\,\,\,\,,\,\,\,\, k_{1}^2=0
\,\,\,\,,\,\,\,\,{\epsilon^\mu}_\mu=0\cr &&\cr
k_2.V^T.\xi\!\!\! &=& \!\!\!0 \,\,\,\,,\,\,\,\,
m_{open}^2=-(2k_2.V^T)^2=0.
\end{eqnarray}
By using these two last equations and
$V^T V={1\over2}(\eta+D_S)$ we find
\begin{equation}
k_1.D_S.k_1=-{1\over2}m_{open}^2=0.\label{mopen}
\end{equation}
By fixing SL(2,R) invariant we find for amplitude
of open-closed scattering from D-brane
\begin{eqnarray}
&&A_{OCD}=g_{c}g_{o}e^{-\lambda}\int d^2z_1dy\langle
V_c V_o\rangle|_{z=i,y=0}\cr &&\cr
&&d^2z_1dy\rightarrow(z_1-y)
(\bar{z}_1-y)(z_1-\bar{z}_1),
\end{eqnarray}
with $(D.V)^{ab}=(V^T)^{ab}$ and $(D.V)^{ij}
=-\eta^{ij}$ after a little computation
we find graviton-gauge field and
graviton-scalar scattering. From this
amplitude we read vertices as
\begin{eqnarray}
(V_{\epsilon_1A})^{a}&\sim&{\kappa c
\sqrt{T_p}\over4}\bigg\{2Tr(\epsilon_1
.D_S)k_1.V_{A}^a-4k_1.D_S.\epsilon_1
.V_{A}^a+4k_1.D_A.\epsilon_1.V_{S}^a
\cr &&\cr\!\!\! \!\!\! &-& \!\!\!\alpha'
\left(k_1.D_S.\epsilon_1.D_S.k_1
+k_1.D_A.\epsilon_1.D_A.k_1\right)
k_1.V_{A}^a\bigg\}\cr &&\cr
(V_{\epsilon_1\lambda})_i&\sim&{\kappa c
\sqrt{T_p}\over4}\bigg\{2Tr(\epsilon_1.
D_S)k_{1i}-4k_1.D_S.\epsilon_{1i}\cr &&\cr
\!\!\! \!\!\! &-& \!\!\!\alpha'\left(k_1.D_S
.\epsilon_1.D_S.k_1+k_1.D_A.\epsilon_1.
D_A.k_1\right)k_{1i}\bigg\},
\end{eqnarray}
which are exactly the same as (\ref{ver})
if one uses the on shell condition for the
states in (\ref{ver}). The last terms in the
vertices in (\ref{ver}) vanish upon using
the massless condition for open strings,
i.e., (\ref{mopen}).

Note that the open string vertex operator
in our calculation corresponds to noncommutative
fields whereas we write the effective
action in terms of commutative fields. The difference
between them is the Siberg-Witten map.
However for the linear open string field
there is no difference between commutative
and noncommutative fields.
\newpage
\pagebreak

\end{document}